# The Formation of Striae within Cometary Dust Tails by a Sublimation-Driven YORP-like Effect

Jordan K. Steckloff[a], Seth A. Jacobson[b,c]

[a]Purdue University, Department of Physics and Astronomy, 525 Northwestern Avenue, West Lafayette, IN 47907

[b]Observatoire de la Côte d'Azur, Laboratoire Lagrange, CS 34429, 06304 Nice Cedex 4, France

[c]Universtät Bayreuth, Bayerisches Geoinstitut, 95440 Bayreuth, Germany

**Abstract**

Sublimating gas molecules scatter off of the surface of an icy body in the same manner as photons (Lambertian Scattering). This means that for every photon-driven body force, there should be a sublimation-driven analogue that affects icy bodies. Thermal photons emitted from the surfaces of asymmetrically shaped bodies in the Solar System generate net torques that change the spin rates of these bodies over time. The long-term averaging of this torque is called the YORP effect. Here we propose a sublimation-driven analogue to the YORP effect (Sublimation-YORP or SYORP), in which sublimating gas molecules emitted from the surfaces of icy bodies in the Solar System also generate net torques on the bodies. However, sublimating gas molecules carry $\sim 10^4$-$10^5$ times more momentum away from the body than thermal photons, resulting in much greater body torques. Previous studies of sublimative torques focused on emissions from highly localized sources on the surfaces of Jupiter Family Comet nuclei, and have therefore required extensive empirical observations to predict the resulting behavior of the body. By contrast, SYORP applies to non-localized emissions across the entire body, which likely dominates sublimation-drive torques on small icy chunks and Dynamically Young Comets outside the Jupiter Family, and can therefore be applied without high-resolution spacecraft observations of their surfaces. Instead, we repurpose the well-tested mathematical machinery of the YORP effect to account for sublimation-driven torques. We show how an SYORP-driven mechanism best matches observations of the rarely observed, Sun-oriented linear features (striae) in the tails of comets, whose formation mechanism has remained enigmatic for decades. The SYORP effect naturally explains why striae tend to be observed between near-perihelion and ~1 AU from the Sun for comets with perihelia less than 0.6 AU, and solves longstanding problems with moving enough material into the cometary tail to form visible striae. We show that the SYORP mechanism can form striae that match the striae of Comet West, estimate the sizes of the stria-forming chunks, and produce a power-law fit to these parent chunks with a power law index of $-1.4^{+0.3}_{-0.6}$. Lastly, we predict potential observables of this SYORP mechanism, which may appear as clouds or material that appear immediately prior to stria formation, or as a faint, wispy dust feature within the dust tail, between the nucleus and the striae.

## 1. Introduction

Linear features sometimes form within the dust tails of "great comets" from the Oort Cloud such as Comet West (C/1975 V1) (Sekanina & Farrell 1978, 1980), Comet Hale-Bopp (C/1995 O1) (Pittichová et al. 1997), Comet McNaught (C/2006 P1), and Comet PANSTARRS (C/2011 L4) (Jones & Battams 2014). These features are generally aligned with either the nucleus of the comet (synchrones) or with the Sun (striae) (e.g. Comet McNaught [C/2006 P1] in Figure 1).  Synchrones are believed to form from ~1-100 μm dust released nearly simultaneously or diurnally from active areas of the comet's surface, which drifts away from the nucleus due to solar radiation pressure (Karchuk & Korsun 2010).  In contrast, the mechanism that creates striae is poorly understood.

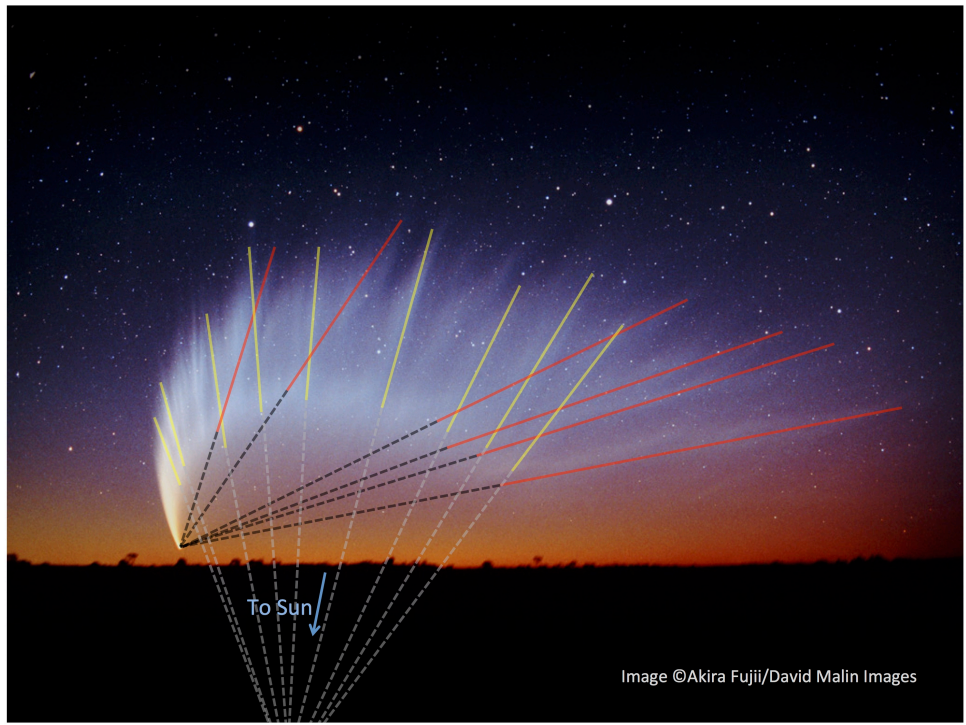

**Figure 1**: *Illustrating Stria and Synchrones.* An image of Comet McNaught (C/2006 P1) shows long linear structures within the tail of the comet.  We have overlain lines to highlight the linear features in the cometary tail.  Note how these features line up with either the head of the comet (synchrones) or with the Sun (striae).  Image ©Akira Fujii/David Malin Images reproduced with permission, with annotations and markings added by authors

Sekanina & Farrell (1980) observed that "striae seem to fit synchronic formations whose sources of emission are located in the area of the dust tail rather than in the nucleus," and postulated three conditions that need to be met by the "parent" materials that form a stria: (1) these materials must be ejected simultaneously from the nucleus; (2) they must experience identical repulsive accelerations from the Sun and (3) these parent objects must break up and disperse simultaneously (listed at the beginning of Section II in Sekanina & Farrell 1980). Some proposed mechanisms assume that (3) occurs as a single, short-lived event (Sekanina & Farrell, 1980; Fröhlich & Notni, 1998), while other mechanisms model (3) as a relatively long-lived fragmentation cascade (Nishioka 1998, Jones & Battams 2014). Regardless of the exact details of (3), these three conditions ensure that the pre-stria materials arrive at the source location of a stria as a single unit, where the parent materials are then transformed into a daughter fragment size distribution that creates the narrow lineaments oriented towards the Sun via anti-sunward acceleration.

## 2. Radiation Pressure

Sekanina & Farrell (1980) and subsequent authors (e.g. Fröhlich & Notni, 1988; Pittichová et al, 1997) considered that solar radiation pressure was solely responsible for the parent materials' repulsive acceleration (second condition above). Sunlight, like gravity, obeys an inverse square law and solar radiation pressure is oriented antiparallel to the solar gravitational acceleration force (to leading order). Thus, its strength can be parameterized by the dimensionless constant β, which is the ratio of the force of solar radiation pressure to the solar gravitational force acting upon a particular object. Since the force of gravity depends on an objects volume (~$R^3$) while force of radiation depends on an objects surface area (~$R^2$), β is a size-dependent parameter. For Comet West, the β parameter for the parent materials released from the nucleus was estimated at $β_p$ = 0.55 – 1.10, while the β parameter for the dust fragments within the striae was $β_f$ = 0.6 – 2.7 (Sekanina & Farrell, 1980). Such high beta parameters require that both parent and daughter grains be small (~ 0.1 μm), such that a small parent grain is most likely capable of creating only ~10 daughter grains (Sekanina & Farrell, 1980). Alternatively, the parent grains could be extremely elongated such that they have a Sun-facing cross-section of a ~0.1 μm grain (Sekanina & Farrell, 1980). Since Comet West's striae are estimated to contain ~$10^6$ kg of material (Sekanina & Farrell 1980), such extreme elongation is unlikely, and more recent research has focused instead on exploring mechanisms that allow a swarm of small-sized parent grains to travel together.

Fröhlich & Notni (1988) propose that such a swarm could travel away from the nucleus in a coherent, optically thick parcel of grains with a narrow range of *β*-values. The breadth of this range depends on the swarm's optical thickness (with optically thin swarms incapable of remaining together), with β values above this range receiving enough illumination to surge ahead and leave the swarm, while grains with β values below this range lag behind the coherent swarm. Fröhlich & Notni (1988) propose that swarms on the order of ~1000 km across become optically thin in the cometary tail and disperse, forming striae. However, to maintain an optically thick swarm the grains must not have any significant transverse velocity (motion perpendicular to the direction of solar gravity/radiation pressure), a condition that is thermodynamically very unlikely without a mechanism for laterally confining the dust.

Neither of these proposed mechanisms is satisfactory. Meeting Sekanina & Farrell's (1980) second condition with radiation pressure requires small parent grains, but then it is difficult to meet the third condition while creating a large enough mass of daughter grains. If an alternative to radiation pressure can be found, then these issues may disappear.

Lastly, observations show that comets with perihelia <6AU form striae between near-perihelion and ~1 AU of the Sun (Pittichová et al, 1997), which suggests that the mechanism driving stria formation must turn off beyond ~1 AU and somehow prevent the formation of observable striae until after the comet has approached the near-perihelion part of its orbit. Since the intensity of solar irradiation decreases smoothly as the inverse-square of heliocentric distance, there is no heliocentric distance at which the solar radiation pressure drops off precipitously. Therefore, if solar radiation pressure drives stria formation, then striae should form at all heliocentric distances, with differences in solar radiation pressure manifesting itself as an increase in the duration of the stria formation process with increasing heliocentric distance.

**3. Sublimation-Driven Stria Formation Model**

In this paper, we propose a sublimation-driven stria formation mechanism that allows for relatively large, volatile-rich chunks of ejected cometary materials to drift into the cometary dust tail and fragment quickly into fine dust, forming cometary dust tail striae. This mechanism also naturally restricts the formation of observable stria until the comet reaches the near- or post-perihelion portion of its orbit and is inactive beyond ~1 AU. We show, through careful consideration of the timescale of stria formation, that this mechanism is consistent with the observed striae of Comet West.

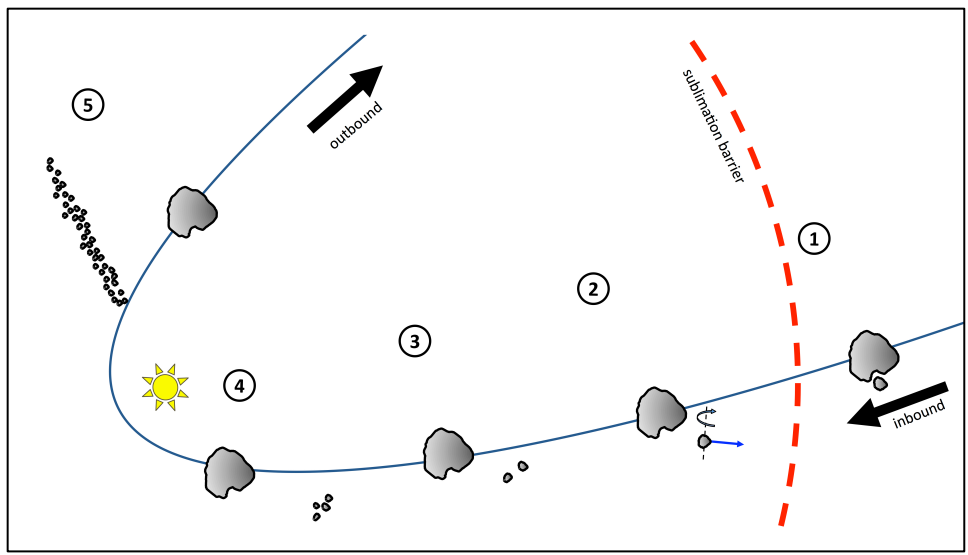

**Figure 2**: *A Cartoon of SYORP-induced Stria Formation.* The five steps of stria formation are illustrated above including (1) parent chunk release, (2) sublimation-driven anti-sunward drift and rotational acceleration, (3) rotational fission, (4) fragmentation cascade, and (5) transition from sublimation to radiation pressure domination of anti-

sunward drift. After step 5, the stream of small micron-sized chunks appears observationally as a stria.

The sublimation of volatile ices is enough to both accelerate the parent chunk anti-sunward relative to the cometary nucleus and spin up the parent chunk to fragmentation, (i.e. rotational fission.) Because the sublimation pressure exerted on the illuminated hemisphere of a volatile rich body is many orders of magnitude greater than radiation pressure, this mechanism is able to affect chunks that are many orders of magnitude larger than previous radiation pressure-driven only mechanisms. We envision that the formation of a stria occurs in five steps (see *Figure 2*): (1) a parent chunk is released from the nucleus of a comet, (2) sublimation pressure causes the parent chunk to drift anti-sunward relative to the nucleus while simultaneously increasing its spin rate, (3) parent chunk spins up to the point of fission, (4) the resulting daughter chunks repeat steps 2 and 3 at an ever-increasing rate, resulting in a fragmentation cascade that (5) stops when the materials become small (micron-sized grains) and devolatilized, at which point radiation pressure dominates the behavior of grains which stream out to form a stria.

Previous studies of the effects of the reactive torques due to sublimating gas on the rotation state of cometary nuclei have focused on the reactive torques from jets either observed or inferred on the surface (e.g. Wilhelm, 1987; Peale & Lissauer, 1989; Julian, 1990; Samarasinha & Belton, 1995; Neishtadt et al, 2002, 2003; Gutiárrez et al., 2003; Sidorenko et al., 2008). These jets may be the dominant rotation state torques for large cometary nuclei (Meech et al., 2011; Belton et al., 2011; Chesley et al., 2013), but the relatively small cometary chunks discussed below are assumed to not possess the ability to create jets (Belton, 2010, 2013; Bruck Syal et al., 2013), although jet production is itself poorly understood. In this work, we propose that it is the background sublimation that torques the cometary chunk. This sublimation is nearly isotropic in the sense that it is emitted from every heated surface element but is very sensitive to the shape and illumination of the chunk. A similar model for an entire comet nuclei has been considered in the past, but it was preliminary (Szegö et al,. 2001), considered only an ellipsoidal shape (Mysen, 2004; 2007), or focused on matching different observational phenomena (Rodionov et al., 2002; Gutiárrez et al., 2007).

*3.1 Step 1: Parent chunks leave comet*

We propose that a single ejected (parent) chunk contains all of the material that later becomes a stria. Sekanina & Farrell (1980) illustrated a method of obtaining an order of magnitude estimate of the volume of a stria for Comet West. Assuming that the dust of a stria has a typical Jupiter Family Comet (JFC) albedo of ~0.03 (Hammel et al. 1987;Brownlee et al. 2004; Lamy et al. 2004; Oberst et al. 2004; Li et al. 2007; Li et al. 2013a; Sierks et al. 2015), is comprised of ~0.1-1 micron particles (Green et al. 2004), and that it originated from an initial parent chunk that was half water ice (McDonnell et al. 1987), then we expect the initial parent chunks to have radii on the order of ~10-100 m. We assume that these parent chunks have a density of ~400 kg/m$^3$, which is typical of JFCs (Sierks et al. 2015; Thomas et al. 2013; Richardson et al. 2007).

Such house- or building-sized (~10-100 m) chunks of material have been observed in the debris of comets 57P/du Toit-Neujmin-Delporte (Fernández, 2009), 73P/Schwassmann-Wachmann 3 (Fuse et al. 2007; Reach et al. 2009), and C/1999 S4 (LINEAR) (Weaver et al. 2001); were observed within the coma of 17P/Holmes following its massive 2007 outburst (Stevenson et al. 2010); and was possibly detected by the *Giotto* spacecraft within a few hundred kilometer of Comet 26P/Grigg-Skjellerup's nucleus (McBride et al. 1997). Most applicably, comet C/1996 B2 (Hyakutake) ejected ~10-100 m chunks, which drifted antisunward relative to the nucleus via sublimation pressure (Desvoivres et al. 2000; Schleicher & Woodney, 2003).

The frequency of striae is likewise consistent with the frequency of ejected ~10-100 m chunks. While a direct measurement of this frequency is difficult due to observational limitations, it is expected to be intermediate to the frequencies of ejection of larger and smaller chunks. Centaur comet 174P/Echeclus ejected a fragment a few kilometers in size (Rousselot, 2008), the only known ejection of such a large fragment. Meanwhile, high-resolution images from spacecraft have revealed that ~1/3 of Jupiter Family Comets (JFCs) eject a large number of decimeter to meter scale chunks into their inner comae at speeds near their escape velocities (~1 m/s) (Hermalyn et al. 2013; Rotundi et al. 2015). Because striae occur more frequently than the ejection of kilometer-scale fragments yet less frequently than the detection of decimeter to meter scale chunks, it is reasonable that the parent bodies that form them are likewise intermediate in size (~10-100 m).

While we do not propose a model for the ejection of these suggested house-sized parent chunks from the nuclei of striated comets, we speculate that perhaps cometary outbursts (Pittichová et al. 1997; Rousselot, 2008) or supervolatile-driven activity may be responsible for launching these parent chunks at greater than escape velocity. Such activity would eject parent chunks with a distribution of initial velocities, and the Rosetta spacecraft observed indirect evidence for the ejection of ~10-100 m chunks from the surface of Comet 67P/Churyumov-Gerasimenko at less than escape velocity that later reimpacted its surface (Thomas et al. 2015). We assume that these parent chunks are rich in water ice throughout, including near the surface of the chunk (relative to the thermal skin depth). If this is not the case, then sublimation pressure will not be able to drive the chunk away from the nucleus (see *Step 2*), due to the inability of the ices to respond to the parent chunk's diurnal thermal cycle.

*3.2 Step 2: Sublimation Pressure instead of Radiation Pressure*

We propose that the reaction force (or equivalently, the sublimative momentum flux) on a volatile-rich parent chunk from the ejection of sublimating gas molecules is enough to both accelerate the parent chunk anti-sunward relative to the cometary nucleus (discussed below) and spin up the parent chunk to fragmentation (discussed in Step 3). Sublimating gasses exert an anti-sunward acceleration on volatile-rich cometary material (Whipple, 1950; Marsden et al. 1973; Steckloff et al.

2015a). Near the Sun, the magnitude of this acceleration behaves similarly to radiation pressure, since it approximates the same inverse square law. Thus, it provides the repulsive acceleration necessary to form striae. However, since the sublimation pressure for $H_2O$ ice is up to 4-5 orders of magnitude stronger than radiation pressure, it can transport chunks of material into the cometary tail that are 4-5 orders of magnitude larger in radius than those transported by radiation pressure alone for a given acceleration of the material relative to the nucleus.

We model parent chunks as balls of pure $H_2O$ ice with such low albedos, that they effectively absorb all incident solar radiation, similar to Steckloff et al. (2015a). We note that these assumptions certainly do no accurately describe the real composition and structure of the parent chunks, which are likely complicated agglomerates of ices and refractory materials with albedos of only a few percent. However, these assumptions illustrate the conditions under which sublimation pressure is maximized, and therefore, define the upper bound of the sublimation pressure acting upon parent chunks. Assuming that the subliming gas is in thermal equilibrium with its source ice and that all incident solar radiation is either re-radiated to space or applied toward overcoming the ice's latent heat of sublimation (Whipple 1950), Steckloff et al. (2015a) show that the sublimation pressure acting on a surface element of cometary material is determined by the following two equations

$$(1-A)\frac{L_{solar}}{4\pi r_h^2 \lambda}\cos\phi = \alpha_{(T)}\sqrt{\frac{m_{mol}}{2\pi RT}}P_{ref}e^{\frac{\lambda}{R}\left(\frac{1}{T_{ref}}-\frac{1}{T}\right)} \quad (1)$$

$$P_{sub\,(r_h,\phi)} = \frac{2}{3}(1-A)\frac{L_{solar}}{4\pi r_h^2 \lambda}\sqrt{\frac{8RT}{\pi m_{mol}}}\cos\phi \quad (2)$$

where $A$ is the bond albedo of the material, $\alpha_{(T)}$ is the temperature-dependent sublimation coefficient of the volatile species, $L_{solar}$ is the Sun's luminosity, $r_{helio}$ is the heliocentric distance of the object, $\lambda$ is the ice's latent heat of sublimation, $\phi$ is the solar phase of the element of surface relative to the subsolar point, $m_{mol}$ is the molar mass of the ice species, $R$ is the ideal gas constant, $T$ is the temperature of the sublimating gas (assumed to be in thermal equilibrium with its source ice), and $P_{ref}$ is an experimentally determined vapor pressure at temperature $T_{ref}$. Since equation (1) is transcendental, we solve for temperature ($T$) numerically, then insert it into equation (2) to determine the sublimation pressure of a given surface area element. This formulation assumes that the coma around the volatile-rich body is optically thin (Steckloff et al. 2015), which is valid for heliocentric distances greater than ~0.05 AU for cometary bodies up to ~1 km (Drahus, 2014). This method of computing sublimation pressures provides similar results to previous methods of computing sublimative forces on comet nuclei (e.g. Whipple, 1950; Marsden et al. 1973; Sekanina, 2003), but is instead based on the theoretical (rather than empirical) relationship between vapor pressure and temperature, and is therefore useful for volatile species for which limited empirical data exists (Steckloff et al. 2015a). We plot this dynamic sublimation pressure at the subsolar point in Figure 3. To compute the net force acting upon a volatile-rich object, we integrate equation 2 over the surface of the object.

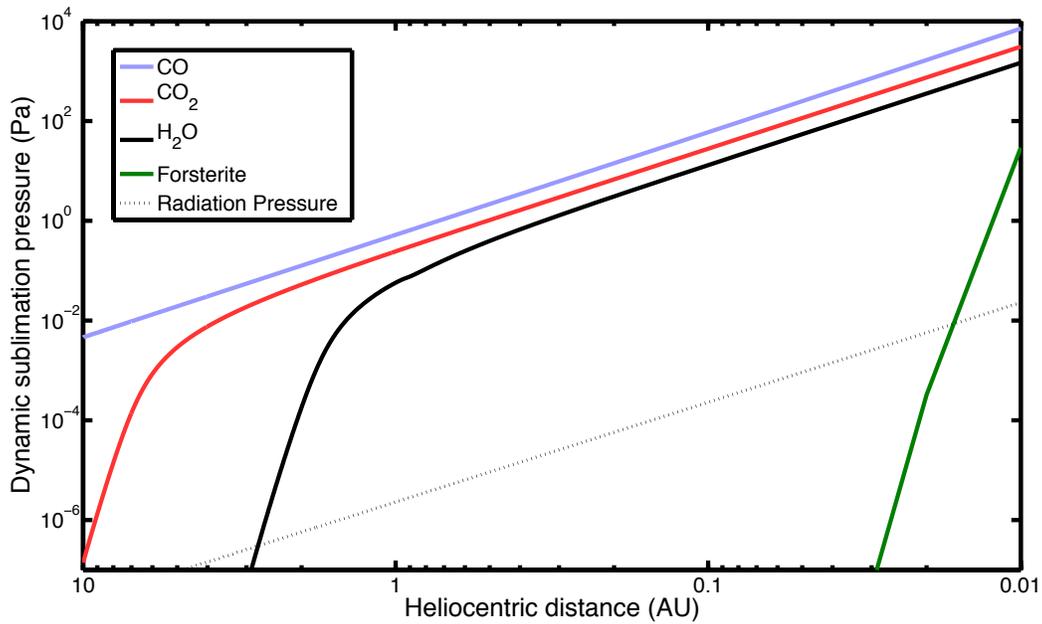

**Figure 3**: *Peak Sublimation Pressure as a Function of Heliocentric Distance.* We adopted *figure 4* from Steckloff et al. (2015a) to show the variation in peak sublimation pressure for an assumed albedo of 0 as a function of heliocentric distance for common cometary volatile species ($H_2O$, $CO_2$, and CO) and the mineral Forsterite, which was found in the coma of comet Wild 2 (Zolensky et al. 2006). For the formation of striae, we focus on the $H_2O$ sublimation curve, as we are positing that $H_2O$ sublimation is responsible for stria formation. Clearly visible is the point (~1 AU) beyond which the sublimation pressure drops off much more quickly. Strength of radiation pressure is added for reference.

Once a parent chunk is broken up into small grains and devolatilized, following the remaining steps detailed below, radiation pressure dominates the non-gravitational behavior of the grains. At this point, radiation pressure streams the chunks into a long lineament as in Sekanina & Farrell (1980), creating the observed striae. However, sublimation pressure is responsible for moving the bulk mass of stria material to the location of stria formation.

*3.3 Step 2 continued: Rotational acceleration due to a sublimation-driven YORP-like Effect (SYORP)*

The back-reaction from anisotropic volatile emission rotationally accelerates striae parent chunks. As a gas molecule escapes from the surface of a parent chunk, it transports angular momentum relative to the center of mass of the parent chunk. The sum of the individual torques from each gas molecule sublimating off of the parent chunk creates a net rotational acceleration of the nucleus (unless that comet possesses perfect symmetry). Thus, in addition to changing the linear motion of a chunk's center of mass, diurnal sublimation can also change a chunk's rotation about its center of mass. We assess the strength of this angular acceleration by analogizing this effect to the well-studied YORP effect (Rubincam, 2000; Bottke et al., 2002; Vokrouhlicky & Capek, 2002; Capek & Vokrouhlicky, 2004; Scheeres, 2007; Rozitis & Green, 2013).

Gas molecules sublimate near the surface of a parent chunk and diffuse through its porous structure, where the gas mean free path is significantly larger than the pores of the cometary material. Eventually these molecules reach the surface, where the last scattering of each gas molecule can be treated independently and the gas emission profile is Lambertian (the probability of being ejected in any given direction is proportional to the cosine of the angle made between that direction and a vector normal to the local surface of the parent chunk [pp. 227-230 in Gombosi, 1994]). Since gas molecules and photons are emitted in a nearly identical fashion, we are able to utilize the theory developed for the photon-driven YORP effect to quantify these sublimation-driven torques.

*3.3.1 The YORP Effect*

Since the numerous instantaneous torques acting on a body are infinitesimal in duration and may be oriented in opposing directions, the YORP effect is a time-averaged phenomenon. The secular rotational acceleration rate due to the YORP effect for an object of radius $R$ and density $\rho$ is (Scheeres, 2007):

$$\frac{d\omega}{dt} = \left( \frac{G_1}{a_\odot^2 \sqrt{1-e_\odot^2}} \right) \frac{3\,C_Y}{4\pi \rho R^2} \tag{3}$$

where $a_\odot$ and $e_\odot$ are the object's heliocentric semi-major axis and eccentricity, $C_Y$ is a shape-dependent coefficient with typical values between $10^{-3}$ and $10^{-2}$ (Scheeres, 2007; Rozitis & Green, 2013), and $G_1 \approx 10^{14}$ kg km s$^{-2}$ is related to the speed of light $c$ and the solar constant $W_\odot = 1.361$ kW m$^{-2}$, which is defined at 1 AU:

$$\frac{G_1}{(1\ \mathrm{AU})^2} = \frac{W_\odot}{c} \tag{4}$$

Note that the magnitude of the rotational acceleration scales inversely with surface area and density, and scales linearly with the absolute strength of the solar radiation pressure at the object's location and with its shape-dependent coefficient $C_Y$, which is defined independent of size (Scheeres, 2007). The coefficient $C_Y$ is determined by the thermally emitted photons, since the absorbed solar radiation contributes no net torque (Rubincam & Paddack, 2010).

*3.3.2 The SYORP Effect*

Since gas molecules carry significantly more momentum than photons, the instantaneous torques acting upon the body are much greater than for the YORP effect. We parameterize this sublimation-driven YORP (SYORP) effect by modifying the YORP effect rotational acceleration equations (equations 3 and 4). Since sublimating gas molecules behave like photons at the surface of the parent chunk, sublimation-driven angular acceleration should depend on the shape of the object in the same manner as emitted photon-driven angular acceleration. Therefore, the shape dependent coefficient for sublimation $C_S$ should be the same as that for photons $C_Y$. Physically, the coefficient $C_S$ represents the fraction of the spin and orbit averaged sublimative momentum flux that contributes a torque due to shape asymmetry. Thus we assume that $C_Y \approx C_S$ for the purposes of our order of magnitude considerations, and should have a value that lies in the range $10^{-3}$ - $10^{-2}$ based on asteroid shapes (Scheeres, 2007; Rozitis & Green, 2013), which should be representative of the shapes of cometary nuclei to first order. This is consistent with recent work that implies the values of $C_S$ for cometary nuclei may lie within a small range of values (Samarasinha & Mueller, 2013).

The absolute strength of the gas sublimation pressure $P_S$ is very different than thermal emission pressure

$$P_Y = G_1/{a_\odot}^2 \sqrt{1-e_\odot^2}. \tag{5}$$

We parameterize this difference with a quantity $\gamma$, which is the ratio of the sublimation pressure to the radiation pressure:

$$\gamma = P_S/P_Y \tag{6}$$

The angular acceleration associated with SYORP is directly analogous to the angular acceleration associated with YORP:

$$\frac{d\omega}{dt} = \frac{3P_S C_S}{4\pi\rho R^2} = \frac{3\gamma P_Y C_Y}{4\pi\rho R^2} \tag{7}$$

where we have taken advantage of both the new parameter $\gamma$ and the equivalence between the two shape factors $C_S$ and $C_Y$.

Since a subliming gas molecule carries significantly more momentum than an emitted thermal photon, we might naively expect $\gamma$ to be greater than one. However, if gas emission is significantly reduced relative to thermal emission, $\gamma$ may be less than one. We use equations 5 and 2 for the radiation $P_Y$ and sublimation $P_S$ pressures respectively to compute the ratio $\gamma$ as a function of heliocentric distance. Near the Sun, the chunk is cooled predominantly through sublimative cooling and energy is lost primarily through overcoming a species latent heat of sublimation. Since the incident solar energy flux scales as the inverse square of heliocentric distance, the sublimative mass-loss rate and resulting sublimation pressures (and therefore gamma) scale approximately (but not exactly) as an inverse square law with heliocentric distance. Further from the Sun, however, the chunk is predominantly cooled by blackbody radiation, and the sublimative mass-loss rates fall far short of the inverse square law, resulting in a steep drop off in gamma with increasing heliocentric distance. This leads to a shape of the gamma curves in which they rise steeply with decreasing heliocentric distance until reaching an approximately constant value (see Figure 4).

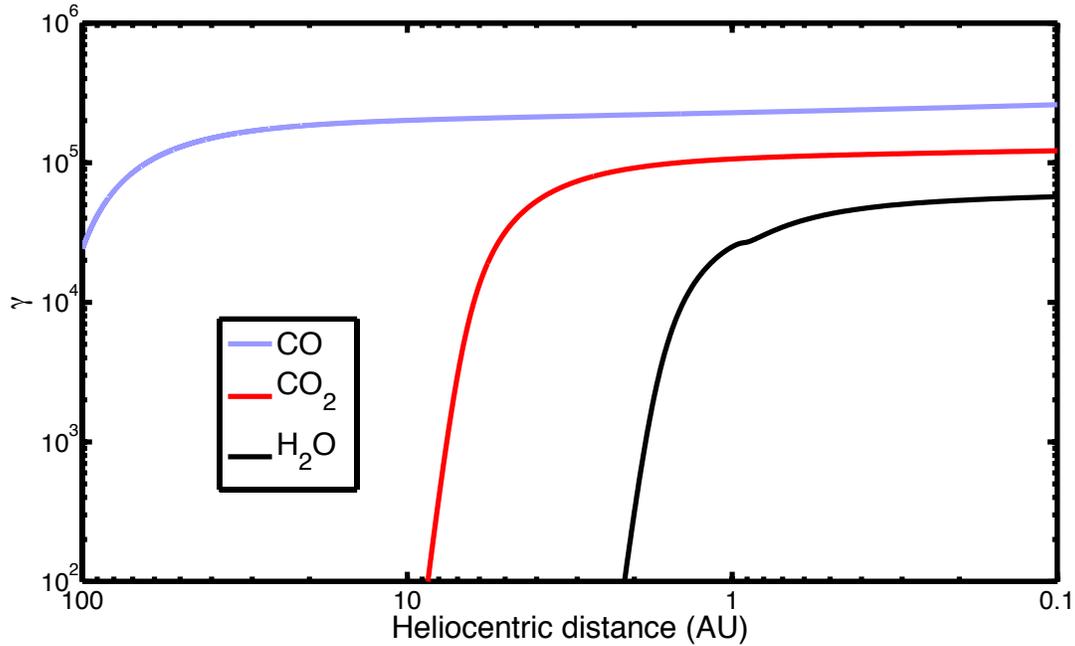

**Figure 4:** *A plot of the gamma factor for various species.* Above is a plot of γ (ratio of sublimation pressure to radiation pressure) versus heliocentric distance for various volatiles. We computed these values based on a planar surface element composed purely of the respective volatile, with the Sun located at the zenith. Sublimation pressure data for all volatiles obtained from Steckloff et al. (2015a). We observe that the volatiles activate at larger heliocentric distances, building up the sublimation pressure as the sublimating object moves inward. Closer to the Sun the volatile becomes fully activated, and nearly scale

with $\propto 1/r_{helio}^2$, causing the γ-gamma curves to flatten out to a nearly a constant value.

*3.4 Step 3: Critical failure of the body*

We define a critical rotation rate $\omega_{crit}$, above which the centripetal acceleration required to hold the body together overcomes the tensile strength of the body, leading to fragmentation. Since these chunks survived ejection from the cometary nucleus intact, they are necessarily stronger than their parent nucleus, which typically have strengths on the order of a few Pascals (Sekanina & Yeomans, 1985; Asphaug & Benz, 1996; Melosh, 2011; Bowling et al. 2014; Steckloff et al. 2015a, Thomas et al. 2015). For these icy chunks, self-gravitational forces are dominated by even this weak strength limit. Thus gravity has a negligible effect in holding these icy parent chunks together. To estimate $\omega_{crit}$, we approximate the icy parent chunks as rectangular prisms, where the long axis ($a = 2R$) is twice the length of the other two sides, which we assume to be equal in length ($b = c = R$). The maximum tensile force exerted along the long axis of the body due to strength is then

$$F_{tensile} = A\sigma_t = R^2 \sigma_t \tag{8}$$

where $A$ is the cross-sectional area perpendicular to the long axis, and $\sigma_t$ is the material tensile strength. The centripetal force at which the body fails (fragments) under principal axis rotation is

$$F_{cent} = ma_{cent} = \tfrac{1}{2}\rho\omega^2 R^4 \tag{9}$$

At the critical rotation rate, $F_{tensile} = F_{cent}$, thus the critical rotation rate (above which the object fragments) is

$$\omega_{crit} = \sqrt{\frac{2\sigma_t}{\rho R^2}} \tag{10}$$

We estimate the SYORP timescale by assuming that the parent chunk starts at rest and compute the amount of time required to spin the chunk up to $\omega_{crit}$. We integrate the expression for angular acceleration (equation 7) with respect to time ($\tau$), set the constant of integration to zero (for chunks starting at rest), and set this resulting expression for angular velocity ($\omega$) equal to $\omega_{crit}$

$$\tau_S = \frac{R\sqrt{32\rho\sigma_t}}{\gamma P_Y C_Y} \tag{11}$$

This timescale defines the duration of an SYORP cycle.

*3.5 Step 4: Runaway Fragmentation Cascade*

We now consider the fragmentation of the parent chunk. Since the chunk slowly spins up to the point of fragmentation, the parent clump likely fragments along a single plane of weakness[1], resulting in two roughly equal-sized daughter chunks. If we assume that the two daughter chunks are equal in mass, and that the total volume of material is preserved, then the daughter chunks will have a radius

---

[1] As opposed to a sudden shock of the material, which may result in many forked fractures and numerous fragments if the shock is traveling faster than the velocity of Raleigh surface waves within the material (order of ~100 m/s)

$\sqrt[3]{1/2}$ of the parent chunk. Such a size decrease is associated with a corresponding increase in the tensile strength of the daughter chunk. According to Griffith Crack Theory (Brace 1961) and assuming a Weibull distribution of flaws within the material, the strength scales approximately as $\sim\sqrt{1/s}$, where $s$ is the size of the object. Thus, the daughter clumps will have a tensile strength that is approximately $\sqrt[6]{2} \approx 1.12$ times the tensile strength of the parent chunk.

After fragmentation, the daughter chunks will be rotating approximately at a rate $\omega_{crit,p}$ (the critical rotation rate of its parent chunk), with the exact value depending on geometry. Thus, instead of starting at rest (as is assumed for the initial parent chunk), the daughter chunks already are rotating at a significant fraction of their own $\omega_{crit}$

$$\omega_0 = \omega_{crit,p} \tag{12}$$

$$= \sqrt{\frac{1}{\sqrt[3]{2^2}\sqrt[6]{2}}}\,\omega_{crit} \tag{13}$$

$$C = \frac{\omega_0}{\omega_{crit}} \approx 0.75 \tag{14}$$

which reduces the time needed for the daughter chunks to spin up to fragmentation proportionally. Therefore, the timescales to fragmentation for all chunks (except for the initial parent chunk) are $(1-C) \approx 25\%$ of the time to rotational fission from rest. Therefore, while the initial parent chunk will require the full $\tau_S$ to spin up to fragmentation, all ensuing daughter chunks will only require $(1-C)\tau_S$ to spin up to $\omega_{crit}$.

If we compute the ratio of the SYORP timescales (equation 11) for the daughter clump versus the parent clump, we find that

$$\frac{\tau_{daughter}}{\tau_{parent}} = \frac{(1-C)R_{daughter}\sqrt{\sigma_{daughter}}}{(1-C)R_{parent}\sqrt{\sigma_{parent}}} \tag{15}$$

$$= \sqrt[3]{1/2}\sqrt[12]{2} \tag{16}$$

$$= \sqrt[12]{\frac{1}{8}} \approx 0.84 \tag{17}$$

assuming that $\rho$ and $C_s$ are the same for parent and daughter chunks. Since this ratio of SYORP timescales is less than 1, each successive generation of chunks will have a shorter lifetime than the previous generation, leading to a runaway cascade of fragmentation. Such a cascade is consistent with the modeling of Nishioka (1998) and Jones & Battams (2014) for the creation of dust necessary to explain striae.

We next estimate the duration of the entire cascade of fragmentation events, which is equivalent to the elapsed time between parent chunk ejection from the nucleus and the onset of stria formation. We first compute the number of fragmentation steps needed to fragment a parent chunk into micron-sized dust, which is the suspected size of stria grains (Sekanina & Farrell, 1980). Since daughter chunks have a radius $1/\sqrt[3]{2}$ times the size of their parent chunks, the radius of a chunk in the $n^{th}$ generation is

$$R_n = R_0 2^{-\frac{n}{3}} \tag{18}$$

where $R_0$ is the size of the initial parent chunk ejected from the nucleus. Thus, the number of generations needed to reach size $R_n$ is

$$n = -3\frac{\log_{10}(R_n) - \log_{10}(R_0)}{\log_{10}(2)} \tag{19}$$

Therefore, a parent chunk of ~10-100 m in radius requires ~70-80 generations to produce micron sized dust.

Since the SYORP timescale decreases with each subsequent generation, we can analytically solve for the total amount of time needed for a parent chunk to fragment into the $n^{th}$ generation

$$T_n = \tau_0 + \tau_0(1-C)\sum_{i=1}^{n}\left(\frac{\tau_n}{\tau_{n-1}}\right)^i \approx \tau_0 + 0.25\tau_0 \sum_{i=1}^{n}(0.84)^i \qquad (20)$$

where $\tau_0$ is the SYORP timescale of the initial parent chunk and $C = \omega_0/\omega_{crit}$, which accounts for the nonzero initial rotation of the daughter chunks. The first ~10 generations, which together reduce parent chunk radii by an order of magnitude, dominate this total timescale, occupying over 90% of the time needed to reach sufficiently small fragments. Thus, the time required for an ejected parent chunk to fragment into micron-sized stria grains (and therefore the duration of the stria-forming fragmentation cascade) is effectively independent of the size of the final grain

$$T_{fragmentation} \approx T_n \approx 2.31\tau_0 \approx 2.31\frac{R_0\sqrt{32\rho\sigma_{t,0}}}{\gamma P_Y C_Y} \qquad (21)$$

where $\sigma_{t,0}$ is the tensile strength of the parent chunk.

After each fragmentation event, classical YORP theory predicts that, on average, half of the daughter chunks will continue to spin up to $\omega_{crit}$, while the other half will spin down towards a stationary state. For those chunks that spin down to a low velocity rotation state, the literature is currently inconclusive as to whether or not they will be captured into a low velocity tumbling state (Vokrouhlický & Capek, 2002; Cicalo & Scheeres, 2010; Breiter et al., 2011). If the chunk is not captured in a tumbling state, then it will pass through a low velocity rotation state and emerge accelerating with the opposite sense of rotation. This has been a standard and successful assumption in the literature matching both near-Earth and main asteroid belt spin period distributions (Rossi et al., 2009; Marzari et al., 2011). After making this assumption, then nominally half the chunks take 175% of the SYORP timescale $\tau_S$ to fragment while the other half take 25% of $\tau_S$. This factor of a few difference of the fragmentation timescale is smaller than the expected order of magnitude variations of the SYORP shape coefficient $C_S$.

When the chunks are large and the SYORP fragmentation timescales are relatively long, the chunks that fragment much faster or much slower than the average chunk could drift away from the pack contributing to background dust production and possibly form separate mini-striae. As the SYORP fragmentation cascade progresses and the fragmentation timescales decrease, even chunks with very different fragmentation timescales will be unable to drift appreciably apart from one another. If only half the initial parent chunk's mass ends up in the stria, then the initial parent chunks must be approximately ¼ larger in radius to account for the mass that fails to form striae. While a sublimative analogue to the Tangential YORP Effect will increase the fraction of chunks that accelerate in the direction of their rotation (Golubov & Krugly, 2012; Golubov et al. 2014) and therefore contribute to stria formation, we conservatively neglect this contribution.

*3.6 Step 5: Onset of Stria Formation*

As the fragmentation cascade continues, the resulting fragments become not only smaller, but also increasingly devolatilized. At some point, the resulting grains within the fragment swarm are so small and devolatilized, that solar radiation pressure dominates their behavior, and they stream anti-sunward as in previous models. While we assume that all daughter chunks are of an equal size and have an idealized distribution of flaws, rotational fragmentation will create chunks that are only approximately equal. While these different sizes will not produce large separations between chunks during earlier generations, variations in size during the final generations will cause the grains to separate from one another via solar radiation pressure according to their differing $\beta$ values, forming a stria (Sekanina & Farrell, 1980). We therefore consider the point at which a parent chunk completes its fragmentation cascade to be the onset of stria formation

*3.7 Modeling and Constraints on Stria Formation*

We now estimate the constraints of SYORP-driven stria formation on Comet West. We approximate Comet West's orbit as a parabola with a perihelion of 0.197 AU, and numerically investigate the heliocentric and cometocentric distances of stria formed from our scheme as a function of the heliocentric distance of parent chunk ejection. We numerically integrate the motion of hypothetical parent chunks ejected from the nucleus between 180 days pre-perihelion to 90 days post-perihelion, and record their heliocentric and cometocentric distances at which they complete their fragmentation cascades. We assume that parent chunks that have not completed their fragmentation cascades by the time they reached a post-perihelion heliocentric distance of 10 AU will not form stria because this distance is much greater than the heliocentric distance beyond which water ice sublimation shuts down. We assume the separation between the comet and the parent chunk is small compared to their heliocentric distances, which allows us to approximate the change in the cometocentric distance ($d_{comet}$) of the parent chunk by assuming that its cometocentric drift is due entirely to the effects of dynamic sublimation pressure

$$\Delta d_{comet} = \tfrac{1}{2} a_{(r_{helio})} \Delta t^2 + vt = \frac{3 P_{sub\,(r_{helio})}}{8 \rho R} \Delta t^2 + v \Delta t \qquad (22)$$

where $a_{(r_{helio})}$ is the acceleration of the parent chunk due to sublimation pressure, $P_{sub\,(r_{helio})}$ is the heliocentric distance dependent sublimation pressure, $\rho$ is the density of the parent chunk, and $v$ is the parent chunk's cometocentric velocity. We assume that this distance montonically increases. This is an admittedly simplified model, which accounts only for a one-dimensional change in the cometocentric distance. However, the largest sources of error are likely the uncertainties in the physical properties of the parent grains. This one-dimensional model is therefore sufficient for our purpose of understanding the order of magnitude behavior of parent chunks, and we reserve two or three-dimensional modeling of stria formation with a deeper study of parent chunk properties for another paper.

Our assumed initial velocity of the parent chunk (~1 m/s) relative to the nucleus is negligible compared to the average velocity needed to move a parent chunk from the nucleus to the cometocentric location of stria formation (~100-1,000 m/s) in the weeks between passing the sublimation barrier (the heliocentric distance within which $H_2O$ sublimation becomes the dominant cooling mechanism of the nucleus) and forming a stria. Therefore, we can treat the parent chunks as though they were initially at rest. Additionally, because the parent chunks have an initial velocity comparable to the comet's escape velocity, the parent chunk will quickly move several nucleus radii away from the nucleus, to a point where the cometary gravity is negligible compared to solar gravity or sublimation pressure (while still being relatively close to the nucleus when compared to the cometocentric distance of stria formation). We therefore ignore the negligible effects of cometary gravity on this calculation. We assume that that parent chunk has a tensile strength of 10 Pa, which is the expected order of magnitude when the ~1 Pa strength of ~1 km comet nuclei (Sekanina & Yeomans, 1985; Asphaug & Benz, 1996; Bowling et al. 2014; Thomas et al. 2015; Steckloff et al. 2015a) is scaled to a ~10 m chunk using a $\sqrt{1/s}$ strength scaling law (Brace, 1961). We use a time step of 6 hours in the numerical modeling.

In Figure 5, we plot the heliocentric distance of the onset of stria formation (the point at which the fragmentation cascade is complete) as a function of the heliocentric distance of ejection of a 10 m parent chunk for a comet with the orbit of Comet West. We plot two different cases of the SYORP coefficient $C_S$, which illustrate two different behaviors in Figure 5: one in which the parent chunk parameters restrict all striae formation to post-perihelion ($C_S$=0.0035), and another in which the parent chunk parameters allow for the formation of some pre-perihelion striae ($C_S$=0.01), which puts a bulge in the curve near perihelion.

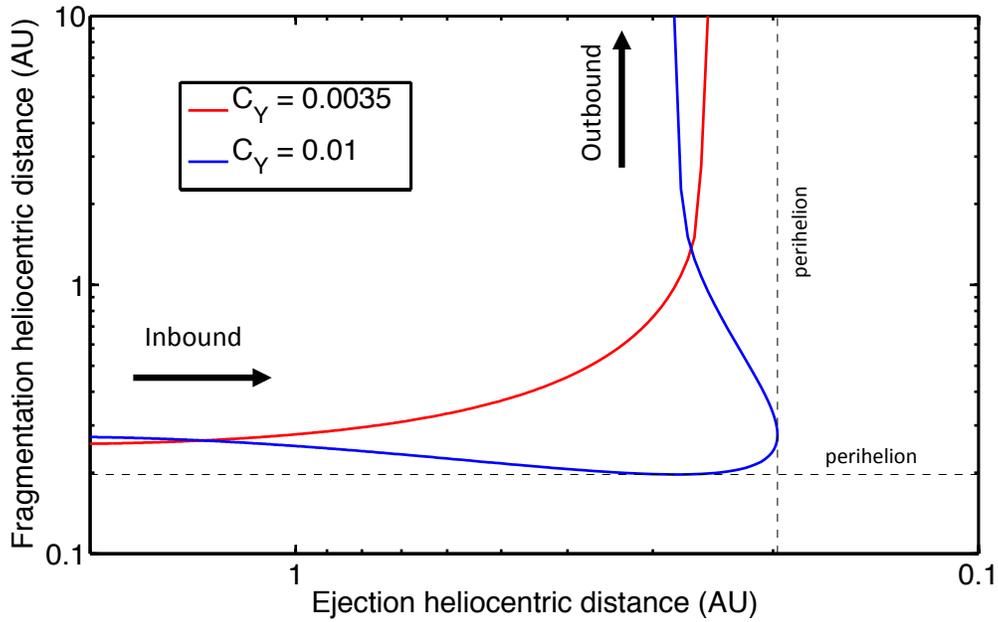

**Figure 5:** *Comet West stria formation heliocentric distance versus parent chunk ejection heliocentric distance.* We plot the heliocentric distance of fragmentation for each simulated 10m parent chunk ejected from Comet West at 6 hour invervals as a function of the heliocentric distance of parent chunk ejection for two values of the SYORP coefficient $C_Y$ = 0.01 and 0.0035. This plot reveals that the overwhelming majority of ejected parent chunks would produce stria between 0.2 and 0.3 AU (near perihelion), consistent with observations (Sekanina & Farrell 1980). Additionally, parent chunks ejected beyond the sublimation barrier (~1 AU) form striae at near the same heliocentric distance (the stria barrier), leading to the asymtotic behavior of the inbound part of the curves. Meanwhile, few chunks ejected after the sublimation barrier have time to fragment

before passing back beyond the sublimation barrier, leading to the asymptotic behavior of the outbound part of the curves.

For the case where $C_S$=0.01 (which is the upper bound of the expected range of SYORP coefficients, and therefore represents the strongest expected response to SYORP), we find that the heliocentric distance of stria formation has little dependence on the heliocentric distance of parent chunk ejection, with the vast majority of parent chunks forming striae within a narrow window of heliocentric distances (for a given parent chunk size and SYORP coefficient). Because of the sublimation barrier, any parent chunk ejected beyond ~1 AU will experience neither a significant SYORP effect nor sublimation pressure until it reaches the sublimation barrier. After crossing the sublimation barrier, the rapid increase in SYORP torques that peak at perihelion will induce a peak in the number of parent chunks completing their fragmentation cascades, and would therefore cause a burst of stria formation near- and post-perihelion. Meanwhile, Figure 5 reveals that very few parent chunks ejected post-perihelion have sufficient time to undergo the SYORP fragmentation cascade (Step 5) to form striae before passing back across the sublimation barrier, and is only possible for parent chunks that have a very strong response to SYORP torques (i.e. smaller radii and larger SYORP coefficients).

Thus, our model predicts that large, Comet West-like stria should preferentially form after the comet reaches near-perihelion and ~1 AU (water sublimation barrier), with a large burst of striae forming near perihelion, consistent with observations of striae (Sekanina & Farrell, 1980; Pittichová et al. 1997). This is not to suggest that no striae can form prior to perihelion. Striated comet nuclei likely eject a population of parent chunks with a distribution of sizes and SYORP coefficients. Because the SYORP response is size-dependent, our model predicts that smaller parent chunks will be able to respond quickly enough to the weaker pre-perihelion SYORP torques to form striae (assuming that SYORP coefficients are independent of size.) However, these early striae would contain significantly less material than the larger striae that form later, and may therefore be unobservable. Thus, while our SYORP model of stria suggests that any comet capable of ejecting icy chunks could form striae, they may not stand out above background dust emission. Therefore, a careful pre-perihelion study of striated comets could confirm this aspect of SYORP theory.

**4. Striae of Comet West**

We lastly apply our model to the striae of Comet West as a proof of concept of the SYORP model. We use this rudimentary one-dimensional model to estimate the sizes and SYORP coefficients of the initial parent chunks needed to match the estimated heliocentric and cometocentric distances of its striae (Sekanina & Farrell, 1980). Sekanina & Farrell (1980) obtain these distances by modeling the motion of devolatilized dust under the effects of solar gravity and radiation pressure. Although the Sekanina & Farrell (1980) model of stria formation differs from the model presented in this paper, both models of dust behavior post-formation are identical. Therefore, the heliocentric and cometocentric distances of stria formation that were obtained by post-formation stria dust modeling are applicable to our model.

We list the heliocentric and cometocentric distances of stria formation for the observed stria of Comet West from Sekanina & Farrell (1980), along with our parent chunk radii and SYORP coefficients ($C_S$) that best fit those distances in Table 1. Each heliocentric and cometocentric distance pair have two unique solutions for parent chunk radius and SYORP coefficient: one solution for the pre-perihelion portion of the comet's orbit, and a second solution for the post-perihelion portion of the orbit. Because the striae in Sekanina & Farrell (1980) were observed post-perihelion, we restrict ourselves to this set of solutions.

| Heliocentric Distance (AU)[1] | Cometocentric Distance (Gm)[1] | Best Fit Parent Radius (m) (error $R^{+10\%}_{-27\%}$) | Best Fit Parent $C_S$ (error $C_S{}^{+6\%}_{-2\%}$) |
|---|---|---|---|
| 0.2284 | 2.56 | 32.5 | 0.00056 |
| 0.2924 | 7.58 | 16.4 | 0.00029 |
| 0.2696 | 5.34 | 20.5 | 0.000355 |
| 0.2581 | 4.2 | 24 | 0.000406 |
| 0.2606 | 4.1 | 24.75 | 0.000415 |
| 0.2535 | 3.27 | 30.5 | 0.000493 |
| 0.2683 | 4.06 | 26.5 | 0.000433 |
| 0.2506 | 2.8 | 35 | 0.000555 |
| 0.2592 | 2.92 | 34 | 0.000530 |
| 0.2517 | 2.14 | 47 | 0.000688 |
| 0.2543 | 1.97 | 50 | 0.000715 |
| 0.2785 | 2.94 | 37 | 0.000544 |
| 0.2769 | 2.29 | 49 | 0.00067 |
| 0.2624 | 1.1 | 95 | 0.00114 |
| 0.2685 | 0.96 | 110 | 0.00126 |
| 0.2841 | 1.12 | 105 | 0.00118 |

[1]Sekanina & Farrell (1980)

**Table 1:** *Heliocentric and Cometocentric locations of stria formation for Comet West and their best-fit parent chunks.* This table lists the modeled heliocentric and cometocentric distances of formation for 16 striae of Comet West (Sekanina & Farrell, 1980). These distances were obtained by modeling the post-formation dynamics of the dust that composed each stria. This table also lists our best fit radius and SYORP coefficient for each stria.

The best-fit parent chunks' SYORP coefficients ($C_S$) lie between 0.00029 – 0.00126, and their best-fit radii lie between 15-110 m. These SYORP coefficients are on the low size of their expected range of ~0.001-0.01 (Scheeres, 2007; Rozitis & Green, 2013), which is based on repurposing YORP coefficients to SYORP. While this may be a result of model assumptions, we acknowledge that it may be indicative of a fundamental difference between the YORP and SYORP effects. The YORP and SYORP coefficients are shape-dependent parameters that describe the second order torques that arise from asymmetries in the shape of the object. Unlike the YORP effect, SYORP depends on the loss of material from the surface of the object that can eliminate asymmetries in its shape over time, particularly at smaller size scales. If the object becomes more symmetrical, its SYORP coefficient will drop over time. Therefore, time-averaged SYORP coefficients may be, as a whole, smaller than their YORP counterparts. While our model assumes a static SYORP coefficient, these best-fit SYORP coefficients are more representative of an average value. Therefore, while the initial SYORP coefficient of a parent chunk may be comparable to its YORP coefficient, the loss of mass required by SYORP may result in a lower average SYORP coefficient than the average YORP coefficient (were the chunk not sublimating.)

The best-fit radii of the parent chunks fall within the expected ~10-100 m range. The estimated error in the size of the radii of the parent chunks is a result of uncertainly in the magnitude of the average dynamic sublimation pressure. Steckloff et al. (2015a) estimate the uncertainly in the sublimation pressure to be up to ~10% for pure $H_2O$ ice sublimation.  Additionally, we use a dynamically new comet C/2012 S1 (ISON), which is a reasonable analogue to the predicted pristine parent chunks, to estimate uncertainties associated with sublimation contributions from less common but more volatile species and the active fraction of the parent chunks' surfaces.  We estimate that the small contributions from less common sublimating volatile species ($CO_2$, CO, etc.) to be up to ~10%, based on their relative abundances (McKay et al. 2014; Weaver et al. 2014) and relative volatilities (Steckloff et al. 2015a).  Unlike JFC nuclei which have only small fractions of their nuclei that are active, the entire surface of Comet ISON appeared to be active (Steckloff et al. 2015b), which is consistent with the thermally primitive nature of long-period comets.  While this would suggest that fragments of such a nucleus (i.e. stria parent chunks) would similarly be active all over, we do not understand what mechanism may be responsible for their ejection.  We consider the case in which the ejection mechanism lofts a partially exposed chunk of material, and conservatively estimate that the exposed region of that chunk (perhaps 20% of its surface) is devolatilized and inactive (or equivalently, that a larger portion of its surface is partially devolatilized).  Because we do not currently have a well-studied ejection mechanism that we could use to better constrain these uncertainties, we adopt the conservative estimate of 20% as the uncertainly in the active area.

These errors are not symmetrical about our best fit solution. Our model assumes the maximum possible sublimation pressure and active area, and uncertainties in their values can only revise them downward. We therefore end up with an asymmetrical error in the average sublimation pressure of $P_{sub}{}^{+10\%}_{-25\%}$ from propagation of errors. We run these uncertainties through our model to estimate the uncertainties in the radii of the parent chunks of the striae from Sekanina & Farrell (1980) to be $\varepsilon R = R^{+10\%}_{-27\%}$, and the uncertainty in the corresponding SYORP coefficients to be $\varepsilon C_S = C_S{}^{+6\%}_{-2\%}$.

With 16 parent chunks, we can generate a Size-Frequency Distribution (SFD), which plots the number of chunks larger than a particular size (see Figure 6). We neglect to include parent clumps smaller than 20 m in this power law fit, as the power law shows a break in the trend, which likely indicates observational bias near the limit of detection. The cumulative-SFD represents the number of chunks greater than a given size, and appears to follow a clear power law ($N(>R) \propto R^q$) with a best-fit power law index ($q$) of -1.4. However, a power law index between -2.0 and -1.1 is consistent with the estimated errors in our model, and power law indexes between -1.1 and -4.0 are consistent with the estimated errors of the chunks up to 50 m in radius. This cumulative-SFD power-law index is consistent with the index of $q$=-1.92±0.20 for Jupiter Family Comets (JFCs) with radii larger than 1.25 km (Snodgrass et al. 2011), but is only marginally consistent with the index of ~-1 that describes the impactor population (<~2km) in the young terrains of Europa (Bierhaus et al. 2012).

The differential Size-Frequency Distribution (differential-SFD) is generated by taking a derivative of the cumulative-SFD with respect to object radius generates the differential Size-Frequency Distribution (differential-SFD).  The differential-SFD for all parent chunks has a power-law slope of -2.4 (-3.0- -2.1), but values between -2.1 and -5 are consistent with the estimated errors of the chunks up to 50 m in radius.  This is consistent with the differential-SFD index of the fragments of Comet 73P/Schwassmann-Wachmann 3 Nucleus B of -2.11 (Fuse et al. 2007) or -2.56 (large fragments F > 10mJy) (Reach et al. 2009), but inconsistent with its small fragments (F < 10 mJy) index of -1.84 (Reach et al. 2009).  This range of differential-SFD power-law indexes for all parent chunks of Comet West is inconsistent with the differential-SFD indexes of -4.7 to -6.6 (Kelley et al. 2013) and -3 to -4 (Rotundi et al. 2015) that describe the chunks and grains in the inner comae of comets 103P/Hartley 2 and 67P Churyumov-Gerasimenko respectively.  However, the two latter populations are for small chunks (up to ~1 m in radius), and the differential-SFD power-law slope for parent chunks of Comet West up to 50 m in radius is consistent with both of these populations.  It is presently unclear whether these similar power law indexes indicate a similar origin, composition, or evolution of these different populations, and further study is warranted to place these cometary populations into a common context and explore how evolutionary and ejection processes may alter these SFDs.

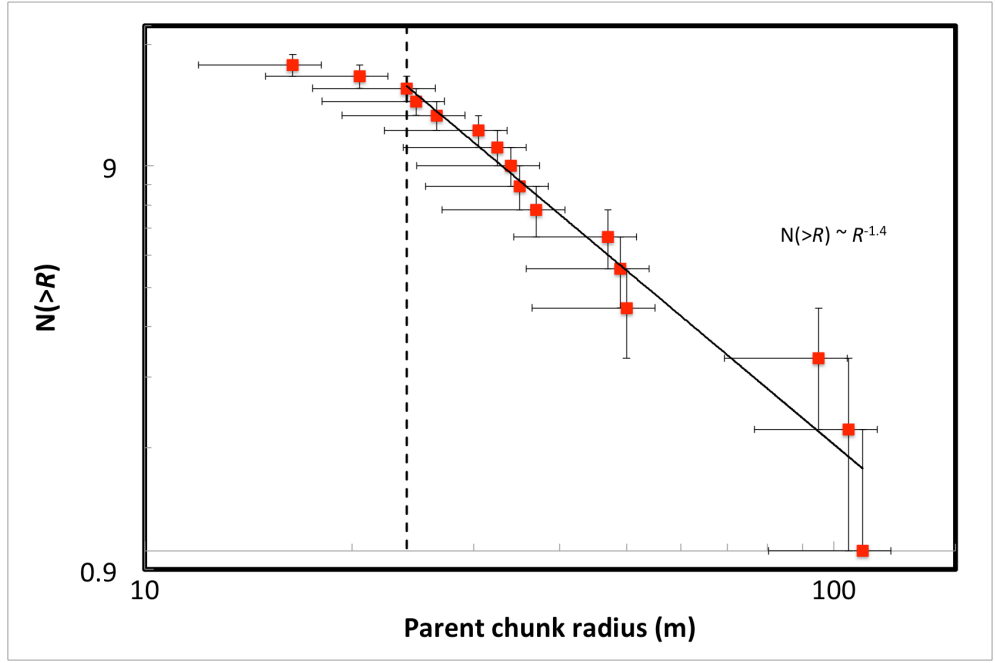

**Figure 6:** *Cumulative Size-frequency distribution of the best-fit parent chunks of Comet West's striae.* Here we plot the number of chunks larger than a given parent chunk size. Vertical error bars are $\sqrt{N}$, while horizontal error bars are the estimated $\sim^{+10\%}_{-27\%}$ uncertainly in parent chunk radius. Vertical dashed line represents a break in the size-frequency distribution, which we belive is due to observational bias.

## 5. Discussion

Thus far our analysis has assumed that the sublimation fronts for the volatile ices are located at the surface of the chunks, rather than below. Comet ISON's dust activity, which is a proxy for gas sublimation, was located predominantly on the sunward side of the nucleus (Li et al. 2013b). This is common for comet nuclei (Whipple, 1950; Keller et al. 1986; Feaga et al. 2007; Belton, 2013; Gulkis et al.

2015), and suggests that the volatile sublimation front is close enough to the surface of the nucleus to respond to the diurnal thermal wave (Steckloff et al. 2015a), such that the time required for a pulse of heat at the surface to propagate to the sublimating volatiles is short compared to the rotation period. This behavior is consistent with the low thermal inertias of cometary material (Lisse et al. 2005; Lamy et al. 2008; Davidsson et al. 2013; Groussin et al. 2013; Gulkis et al. 2015). However, if the rotation period of a parent chunk were to become comparable to this thermal lag time during SYORP spin up, then the chunk's gas emissions would begin to lose their sunward directionality, and sublimation pressure would begin to cease driving the chunk anti-sunward.

Shutting down the anti-sunward sublimation-driven acceleration would not affect the SYORP torques, which, like the YORP effect, only depends on the shape of the chunk. Therefore a chunk in this situation would cease to accelerate heliocentrically, but would drift cometocentrically at a constant rate and continue to spin up to the point of fragmentation, at which point this cycle would repeat with the daughter chunks. Because the antisunward acceleration would episodically shut down, the resulting cometocentric distance of stria formation would be reduced. However, since neither the thermal lag time between the nuclear surface and the volatile ices nor the depth of the volatile ices of Oort Cloud comets is known, these considerations are currently merely unconstrained speculation.

The SYORP mechanism, while explaining why most observed striae form near- or post-perihelion, predicts that striae may also form pre-perihelion within ~1 AU of the Sun. However, the parent chunks that would form these earlier striae would have to undergo their fragmentation cascades in a shorter period of time, and would therefore be significantly smaller than the parent chunks that form post-perihelion striae. Because these smaller parent chunks would form striae that contain less material than the post-perihelion striae, these earlier striae are expected to be faint and likely to remain undetected. A careful pre-perihelion study of comets that produce post-perihelion striae may be able to confirm this aspect of the SYORP theory.

Additionally, we assume that $H_2O$ sublimation is driving the stria formation process. However, if more volatile species such as $CO_2$ or CO are driving striae formation, then striae may form further from the Sun, form faster, and contain more material. Additionally, if parent chunks are ejected via sublimation of supervolatile species from a discrete location of the nucleus, then parent chunks may be diurnally ejected. If this process occurs within the sublimation barrier of the driving species, then it may lead to the formation of striae that are regularly spaced within the cometary tail, and that form at an interval approximating the rotation period of the nucleus.

Our model relies on the ability of comet nuclei to eject ~10-100 m sized chunks at escape velocity (~1 m/s). Long-period comet C/1992 B2 (Hyakutake) experienced an outburst that ejected chunks consistent with the parent chunks in our model (Desvoivres et al. 2000; Schleicher & Woodney, 2003). However, it is unclear whether or not the comet formed striae due to limited observations of the comet post-perihelion. Similarly, Jupiter Family Comet 17P/Holmes produced fragments consistent with parent chunks (Stevenson et al. 2010), however its distant perihelion of 2 AU would likely prevent the vigorous sublimation that is necessary in our model to form striae.  Spacecraft flybys of comet nuclei (such as Giotto, Deep Space 1, Deep Impact, Stardust, DIXI, and Stardust-NExT) would be very unlikely to resolve the ejection of parent-sized chunks of material due to their limited time of encounter, and would almost certainly require a Rosetta-style mission to observe the nucleus of a striated comet for an extended period of time.

The Rosetta spacecraft itself has observed decimeter to meter-sized chunks of material moving at near escape velocity at Comet 67P/Churyumov-Gerasimenko (Rotundi et al. 2015) and would certainly be able to detect the ejection of objects as large as parent chunks.  However, because striae are a rare phenomenon and Jupiter Family Comets are so thermally processed, we would not necessarily expect that 67P/Churyumov-Gerasimenko would be able to eject parent chunks at escape velocity, which is required to form striae.  Indeed, Rosetta has discovered ~10-100 m chunks of material that may have been ejected from the nucleus of 67P/Churyumov-Gerasimenko, but lacked sufficient velocity to escape the nucleus' gravity (Thomas et al. 2015).  Direct observation of the ejection of ~10-100 m chunks of material would be much more likely by a spacecraft at a long period comet or active centaur.  However, failure to detect the ejection of ~10-100 m chunks of material at these bodies would not necessarily invalidate this theory, since it predicts that only some bodies are capable of ejecting these chunks.

It is plausible that a particularly active comet could eject parent chunks at velocities an order or two of magnitude greater than the comet's escape velocity. Such parent chunks could drift significantly farther from the nucleus than other parent chunks, and would form striae far from the cometary tail. However, if these parent chunks are ejected sufficiently far from the Sun in the centaur region, they may drift so far from the nucleus that they would form dust features too far from the nucleus to be easily associated with the comet. The Rosetta spacecraft currently in orbit around the nucleus of comet 67P/Churyumov-Gerasimenko may be able to directly observe the ejection of large chunks of material from the nucleus during perihelion, and perhaps even obtain a velocity profile of the ejected population.

Additionally, our model may predict observable intermediate stages of stria formation. Because we begin with a single parent chunk, the initial stages of stria formation would be unobservable. We have already shown that daughter chunks with radii that are comparable to the initial parent chunk predominantly occupy the duration of the SYORP fragmentation cascade. Thus, as the daughter chunks drift away from the nucleus, they remain unobservable. However, as the runaway fragmentation cascade nears completion, a very large number of small chunks are produced very rapidly. Thus, immediately prior to the onset of stria formation, an observable cloud of material may appear in the tail of the comet that then streams outward into a stria.

While we assume that each step of the SYORP fragmentation cascade produces two identical daughter chunks (size and shape), it is likely that these two chunks vary from one another. If this variation is small, then the fragmentation cascade would be insignificantly affected and the daughter chunks will still complete their fragmentation cascades at approximately the same time. However, if this variation is large, then one daughter chunk may undergo a significantly faster fragmentation cascade, and complete its fragmentation before drifting a significant distance from the nucleus. This would manifest itself as a source of fine-grained debris in the tail of the comet located between the nucleus and the striae. Additionally, if the fragmentation of the larger daughter chunks (early stages of the fragmentation cascade) is messy and produces fine-grained debris, then it would also manifest itself as an additional source of fine-grained material between the nucleus and striae. In either of these cases, one may see a diffuse or wispy tail of material distinct from the striae or the rest of the dust tail. However, if the fragmentation cascade is more ideal, or if the dust tail is bright, then this feature may not be visible or even existent.

Lastly, while we have only applied SYORP to parent chunks on the order of ~10-100 m in radius, there is no reason why SYORP would not affect much larger icy objects within the Solar System. Indeed, the SYORP mechanism should be able to change the spin state of icy objects of all sizes. The limiting factor for SYORP is heliocentric distance, as the effect shuts down beyond the sublimation barrier of the driving volatile species. While we have here only considered the sublimation of water ice (which shuts down beyond ~1 AU), $CO_2$-driven SYORP would be active out to ~10 AU, while CO-driven SYORP would remain active out to ~100 AU! Therefore, as long as the appropriate volatile species is present and abundant, SYORP can provide torques to objects throughout the observable Solar System.

## 6. Summary & Conclusions

We have proposed a new sublimation-driven model for the formation of striae within the tails of comets that provides a natural explanation for why comets with perihelia within 0.6 AU only form striae within ~1 AU of the Sun after reaching the near-perihelion portion of their orbits. Our model easily allows a large amount of material to be transported as a single unit to the location of stria formation, a major weakness of existing stria formation schemes. As part of our driving mechanism, we describe a new, sublimation-driven analogue to the YORP effect (SYORP), which allows large (~10-100 m) parent chunks to fragment quickly enough to form stria within the inner Solar System. If large numbers of parent chunks with similar sizes and shapes are ejected prior to the comet passing within the sublimation barrier, then these parent chunks should produce a sudden burst of striae. However, the ejection of parent chunks with a range of sizes and shapes is more likely.

We apply our model to the striae of Comet West, and find that parent chunks with radii between 15 m and 110 m ($^{+10\%}_{-27\%}$), which are consistent with expected sizes. The sizes of these parent chunks follow a power-law cumulative size-frequency distribution (cumulative-SFD) with a power-law index of $-1.4^{+0.3}_{-0.6}$ ($-1.5^{+0.4}_{-2.5}$ for parent chunks less than 50 m radius), which is consistent with the index of -1.92±0.20 for Jupiter Family Comets with radii larger than 1.25 km (Snodgrass et al. 2011) and marginally consistent with the index of ~-1 for the impactor population into the young terrains of Europa (Bierhaus et al. 2012). The differential Size-Frequency Distribution (differential-SFD) of $-2.4^{+0.3}_{-0.6}$ is consistent with 73P/Schwassmann-Wachmann 3 Nucleus B's large fragments (Reach et al.

2009) or all fragments (Fuse et al. 2007). The differential-SFD for parent chunks less than 50 m in radius of $-2.5^{+0.4}_{-2.5}$ is consistent with the differential-SFD indexes of the particles in the inner comae of comets 103P/Hartley 2 (Kelley et al. 2013) and 67P/Churyumov-Gerasimenko (Rotundi et al. 2015). The mechanism responsible for lofting these parent chunks off of the surface of the nucleus is unknown, but we speculate that it may be the resulting gas drag from a cometary outburst, consistent with the observed parent-sized chunks of comet 17P/Holmes (Stevenson et al. 2010) and comet C/1996 B2 (Hyakutake) (Desvoivres et al. 2000; Schleicher & Woodney, 2003). The SYORP coefficients ($C_S$) of Comet West's parent chunks are 0.00029 – 0.00126 ($^{+6\%}_{-2\%}$), which is on the low side of the expected range of ~0.001-0.01 (Scheeres, 2007; Rozitis & Green, 2013). This may be due to the loss of surface material that is inherent in the SYORP mechanism, and which may remove the asymmetries in the shape of the body that generate the sublimative torques that create the SYORP effect.

We also predict that fainter, potentially observable striae may form earlier than the larger easily observable striae. However, these early striae would tend to form from smaller parent chunks, and would therefore be harder to detect. Additionally, our mechanism suggests that any comet capable of ejecting icy chunks can produce striae, which may or may not be large enough to be observable. Lastly, we speculate on possible intermediate stages of stria formation in our mechanism that may be observable. One would appear as a cloud of material present immediately prior to stria formation, which may or may not be visible above the background of the dust tail. The other depends on imperfections during the SYORP fragmentation cascade, and may appear as a faint wispy tail-like feature located in the dust tail between the nucleus and the striae if the fragmentation is sufficiently imperfect and the dust tail is sufficiently dim.

## 7. Acknowledgements

We wish to thank and acknowledge H. Jay Melosh (Purdue University), Michael Combi (University of Michigan) and Nalin Samarasinha (Planetary Science Institute) for useful conversations and comments that improved this project. Jordan K. Steckloff wishes to thank H. Jay Melosh (Purdue University) for funding this research and providing helpful comments and context. Seth A. Jacobson wishes to acknowledge support from the European Research Council (ERC) Advanced Grant "ACCRETE" (contract number 290568).